\documentclass[sigconf]{acmart}

\usepackage{balance}
\usepackage{tabularx}
\usepackage{tabularx}
\usepackage{multirow}
\usepackage{diagbox}
\usepackage{color}
\def\BibTeX{{\rm B\kern-.05em{\sc i\kern-.025em b}\kern-.08emT\kern-.1667em\lower.7ex\hbox{E}\kern-.125emX}}

\setcopyright{acmcopyright}
    
\newcommand{\tabincell}[2]{\begin{tabular}{@{}#1@{}}#2\end{tabular}}

\begin{document}
\fancyhead{}
\copyrightyear{2019}
\acmYear{2019}
\acmConference[CIKM '19]{The 28th ACM International Conference on Information and Knowledge Management}{November 3--7, 2019}{Beijing, China}
\acmBooktitle{The 28th ACM International Conference on Information and Knowledge Management (CIKM '19), November 3--7, 2019, Beijing, China}
\acmPrice{15.00}
\acmDOI{10.1145/3357384.3358038}
\acmISBN{978-1-4503-6976-3/19/11}
%
\title{GRAPHENE: A Precise Biomedical Literature Retrieval Engine with Graph Augmented Deep Learning and \\External Knowledge Empowerment}

\author{Sendong Zhao, Chang Su, Andrea Sboner, Fei Wang}
\affiliation{%
  \institution{Weill Cornell Medical College, Cornell University}
  \streetaddress{Weill Cornell Medicine}
  \city{New York}
  \country{USA}
  }
\email{{sez4001, csu4001, ans2077, few2001}@med.cornell.edu}


%
\renewcommand{\shortauthors}{Zhao et al.}

%
\begin{abstract}
Effective biomedical literature retrieval (BLR) plays a central role in precision medicine informatics. In this paper, we propose GRAPHENE, which is a deep learning based framework for precise BLR. GRAPHENE consists of three main different modules 1) graph-augmented document representation learning; 2) query expansion and representation learning and 3) learning to rank biomedical articles. The graph-augmented document representation learning module constructs a document-concept graph containing biomedical concept nodes and document nodes so that global biomedical related concept from external knowledge source can be captured, which is further connected to a BiLSTM so both local and global topics can be explored. Query expansion and representation learning module expands the query with abbreviations and different names, and then builds a CNN-based model to convolve the expanded query and obtain a vector representation for each query. Learning to rank minimizes a ranking loss between biomedical articles with the query to learn the retrieval function. Experimental results on applying our system to TREC Precision Medicine track data are provided to demonstrate its effectiveness. 
\end{abstract}

%
%

\ccsdesc[500]{Information systems~Information retrieval}
\ccsdesc[500]{Information systems~Document representation}
\ccsdesc[500]{Information systems~Learning to rank}
\ccsdesc[300]{Computing methodologies~Neural networks}
\ccsdesc[300]{Applied computing~Life and medical sciences}

%
\keywords{Biomedical Literature Retrieval, Graph Augmented Document Representation Learning, Deep Neural Networks, Learning to Rank}

%

%
\maketitle

\section{Introduction}

Precision medicine \cite{collins2015new}, with the goal of providing the right treatment to the right patient at the right time, holds great premise on treating complicated diseases such as cancer. Biomedical literature database stores the state-of-the-art information about the various aspects related to precision medicine (e.g., patients, diseases, genetics, treatments, etc.) and serves as a vital knowledge source. For example, with the prevalence of Next Generation Sequencing (NGS) technology, vast amounts of genetic variants can be identified for a specific patient on a specific tissue (e.g., tumor). The domain experts, such as clinicians, pathologists, oncologists, need to search the relevant literature on PubMed to interpret these genetic variants. Therefore, an efficient and accurate biomedical literature retrieval (BLR) system is crucial.

There are many challenges to building such an effective BLR system. For example, the vocabulary of professional terms in biomedical articles is typically large,
and there are many semantic variations for the same biomedical concept (e.g., a specific genetic variant), the relations between these concepts are complicated (which could be explicit/implicit, direct/indirect, and known/unknown, etc.). 

The goal of this paper is to develop a high-performance BLR system in view of the above challenges.  
Specifically, our task is to take patient information with the genetic variant as query and retrieve relevant biomedical articles from literature databases such as MEDLINE\footnote{MEDLINE (\url{https://www.nlm.nih.gov/bsd/medline.html}) is a bibliographic database of life sciences and biomedical information. It includes bibliographic information for articles from academic journals covering medicine, nursing, pharmacy, dentistry, veterinary medicine, and health care.}. 
Here the patient information can be either in a structured form (disease name, variant, demographic, ...) or unstructured text. The output should be a list of biomedical articles which are ranked according to the relevance with the query.

\begin{figure*}
	\centering
	\includegraphics[width=0.98\textwidth]{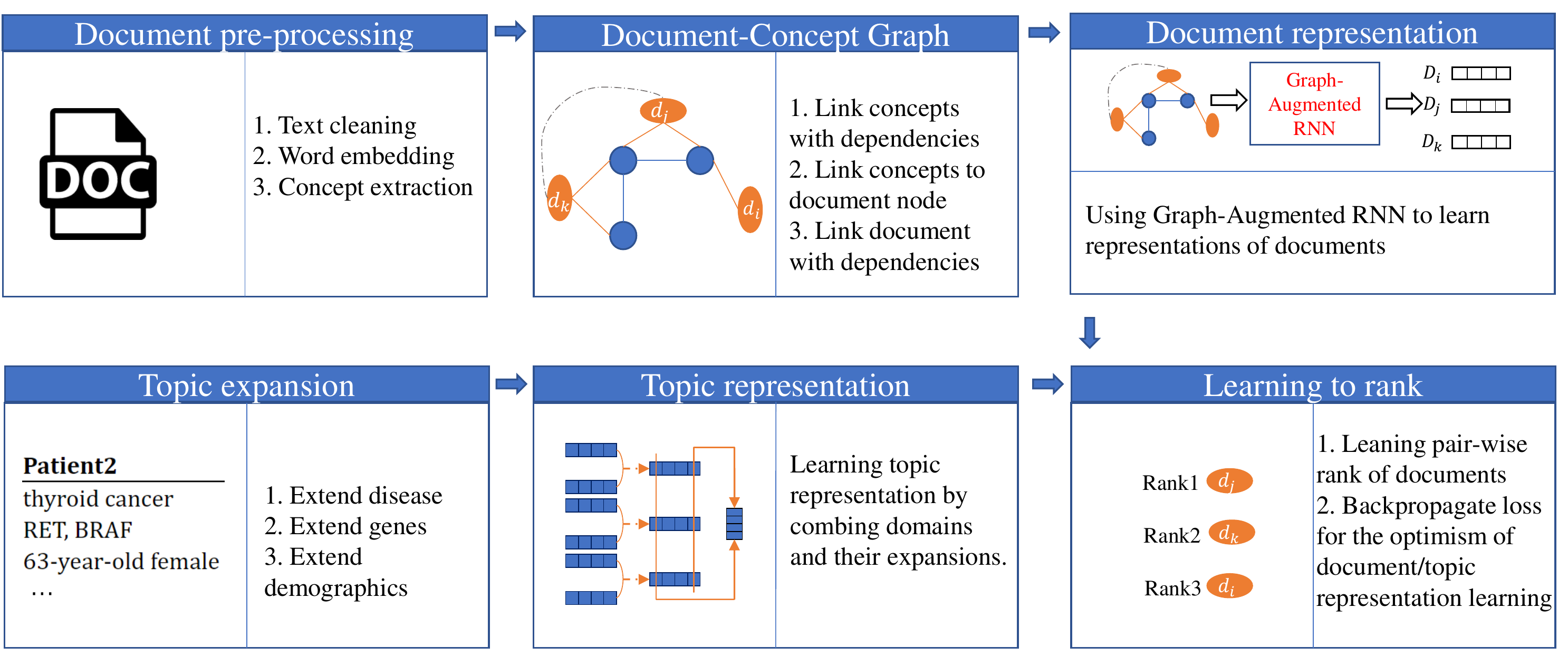}
	\caption{The framework of graph augmented deep learning with knowledge empowerment engine for for biomedical literature retrieval.}
	\label{fig:frameword}
\end{figure*}
There are mainly two types of retrieval models can be used for our scenario. The first type is the traditional document retrieval model which represents the query and documents both as one-hot representations and matches query and documents with similarity measures such as cosine similarity. In particular, one-hot representations can be generated through the bag-of-words model or TF-IDF model. In the bag-of-words model, a text (such as a query or a biomedical article) is represented as the bag (multiset) of its words, disregarding grammar and even word order but keeping multiplicity. TF-IDF model is a better approach compared to the bag-of-words model. It assigns a weighting to each word in the document and put that weighting in the vector, which is allowed to normalize the count/frequency of the word. Yet the drawback of the traditional retrieval model is apparent. Consider one takes "lung cancer" as the query, then the model can only retrieve the articles with exactly the term "lung cancer" in their contents. A notable exception is PubMed which expands the query by mapping it to related MeSH terms \cite{lu2009evaluation}. Although this increases recall, it often decreases precision which is crucial for precision medicine \cite{hersh2000assessing}. 
The second type is deep learning models, such as DeepMatch \cite{lu2013deep} and Delta \cite{mohan2018fast}, which have demonstrated superior performance in recent years. The advantage of these models comes from their flexible representation learning process for the query and the documents. In particular, the learned representations can encode the underlying semantics for queries and documents. This greatly enhanced the capturing of the semantically similar words but with different expressions such as "cancer" and "tumor", "diabetes" and "hyperglycemia". Therefore, take "diabetes" as query these models might find articles with "hyperglycemia". However, it is still difficult for these models to identify indirectly/implicitly related terms such as "diabetes" and "metformin", in which case we would need external knowledge base \cite{zhao2018causaltriad}. 


Our developed system is called GRAPHENE, which stands for \textbf{GR}aph \textbf{A}ugmented dee\textbf{P} learning wit\textbf{H} knowl\textbf{E}dge empowerme\textbf{N}t \textbf{E}ngine.  
Our model learns literature representations with local text and external structured knowledge and matches queries with the pair-wise learning to rank mechanism. The overall architecture of GRAPHENE is shown in Figure~\ref{fig:frameword}, which is composed of 3 main modules. The first is a document representation learning module. We propose to construct a document-concept graph, upon which a graph-augmented document representation is learned to encode both the underlying semantics of the literature and the information from non-local relevant medical concepts. The second is patient information query representation learning module, where we propose to expand query and learn representations through the convolution upon the expanded query. The third is a learning to rank module, which minimizes the pairwise ranking loss between the patient information and biomedical article to learn a partial order of relevance of biomedical articles. The organic integration of these components makes GRAPHENE possible to retrieve biomedical articles effectively and precisely.

Experimental results on TREC Precision Medicine data \cite{roberts2017overview} demonstrate that our model can effectively retrieve most relevant biomedical articles with patient information. 
The results also show that our model outperforms those deep neural retrieval models which represent documents with textual information only, suggesting the necessity of encoding graph-based external knowledge into document representation. 

The paper is organized as follows. Section 2 includes related studies in biomedical literature retrieval and document retrieval methodology. Section 3 describes the details of graph-augmented document representation learning algorithm. Section 4 describes the details of CNN-based query representation learning algorithm and pair-wise learning-to-rank algorithm.  Experimental results are presented in
Section 5, and the paper is concluded in Section 6.

\section{Related work}
\subsection{Biomedical Literature Retrieval}
There has already been a lot of research on biomedical literature retrieval \cite{fiorini2018best,fiorini2017cutting,fiorini2018user,mohan2018fast,Soldaini2017cikm,Zheng2016graph}. 
For example, Zheng and Wan \cite{Zheng2016graph} proposed to use the paragraph vector technique to learn the latent semantic representation of texts and treat the latent semantic representations and the original bag-of-words representations as two different modalities. They then proposed to use the multi-modality learning algorithm to retrieve biomedical literature for clinical decision support. Soldaini et al. \cite{Soldaini2017cikm} applied query reformulation techniques to address the need of literature search based on case reports. Best Match \cite{fiorini2018best} is the first relevance search algorithm for PubMed that leverages the intelligence of users and machine-learning technology as an alternative to the traditional sorting techniques. 
Delta \cite{mohan2018fast} is a deep learning based model that applies convolution operation upon an updated document matrix in which each word is replaced with the most similar word in the query. However, this model takes the key topic word and other words in query with equal importance, making the retrieval out of focus. 

\subsection{Document Retrieval Models}
The traditional document retrieval models represent both the queries and documents as one-hot representations and match query and documents with similarity measures like cosine similarity. As we stated in the introduction, the problem of this method is that the semantically similar query-document pairs without exact expression match cannot be identified. 

In recent years, deep learning based retrieval models, such as DeepMatch \cite{lu2013deep},  PACRR \cite{hui2017pacrr}, Delta \cite{mohan2018fast},  Conv-KNRM \cite{Dai2018WSDM}, MASH RNN \cite{jiang2019semantic}, have dominated the document retrieval research. In particular, these models have been focusing on 1) flexible representation learning for query and documents and 2) measuring the similarity between query and documents at different levels. Correspondingly, there are two main categories of deep neural information retrieval (IR) models. One is the representation-focused model, which tries to learn good representations for both query and documents with deep neural networks, and then conducts matching between the learned representations. Examples include DSSM \cite{huang2013learning}, C-DSSM \cite{gao2014modeling}, ARC-I \cite{hu2014convolutional}, Delta \cite{mohan2018fast}, MASH RNN \cite{jiang2019semantic}. The other is the interaction-focused model, which first builds local interactions (i.e., local matching signals) between the query and documents, and then uses deep neural networks to learn the overall matching score. Examples include DeepMatch \cite{lu2013deep}, ARC-II \cite{hu2014convolutional}, DRMM \cite{guo2016deep}, ESR \cite{xiong2017explicit}, PACRR \cite{hui2017pacrr}, Conv-KNRM \cite{Dai2018WSDM} and SMASH RNN \cite{jiang2019semantic}.

These deep neural IR models can effectively exploit the underlying semantics for queries and documents, which can be good at matching similar but differently expressed words like "cancer" and "tumor", "diabetes" and "hyperglycemia". Therefore, take "diabetes" as query the model might find articles with "hyperglycemia". But these model cannot find articles indirectly/implicitly related to "diabetes" like the article with "metformin" because it requires external knowledge (metformin, against, diabetes).
Although prior research generally confirms that external knowledge has great value for document retrieval \cite{scells2017integrating,goodwin2016medical,xiong2015esdrank}, little research has been conducted to show the value of the external knowledge on deep neural IR models. There are studies trying to perform query expansion or leverage pre-trained embeddings of name entities using external knowledge bases. For example, Xiong and Callan \cite{Xiong2015} presented a simple and effective method to improve query expansion with Freebase. They proposed a supervised model to combine the information derived from Freebase descriptions and categories so that `effective terms can be selected for query expansion. In another paper, Xiong et al. \cite{xiong2017explicit} proposed a ranking technique for matching query and documents, where a knowledge graph is leveraged to pre-train the embeddings of the named entities in both query and documents.

\begin{table}[tb]
	\caption{Symbols and descriptions.}
	\vspace{-0.1in}
	\label{tab:notation}
	\centering
	\begin{tabularx}{0.47\textwidth}{|l|X|}
		\hline
		Symbol & Description\\
		\hline
		$q_k$ & $k$-th query with patient information or MeSH terms\\
		\hline
		${d_i}$ & document nodes, each of them consists of a sequence of $k$ words $(x_1^{(i)}, x_2^{(i)}, ..., x_k^{(i)})$\\
		\hline
		$d_{p_k}^{(j)}$& the document which is ranked as $j$-th in the ranking list for query $p_k$\\
		\hline
		${c_j}$ & biomedical concept nodes\\
		\hline
		$\mathbf{h}^{(i)}_k$ & the $k$-th hidden state of RNN in modelling document $d_i$\\
		\hline
		$\mathbf{g}_{d_i}^{(l)}$& the hidden state of document node $v_{d_i}$ in the $l$-th layer of the graph neural network.\\
		\hline
		$\mathbf{g}_{c_i}^{(l)}$& the hidden state of biomedical node $v_{c_i}$ in the $l$-th layer of the graph neural network.\\
		\hline
	\end{tabularx}
\end{table}


\section{Representation Learning for Document-Concept Graph}

In this section, we present a graph-augmented representation learning approach for documents and biomedical concepts through the Document-Concept Graph (DCG). There are two types of nodes in a DCG: the biomedical concepts and documents. There are also two types of edges in the graph. One links concepts and documents according to their co-occurrence relationships. The other links pairwise concepts if any semantic relationship can be identified between them in a specific knowledge base. 
With the DCG representation, the global topics of the documents are effectively encoded through the document-concept relationships. Moreover, such topics can further be enhanced with external knowledge sources through the transitions on the graph.
In this paper, we denote scalars by lowercase letters, such as $x$; vectors by boldface lowercase letters, such as $\mathbf{x}$; and matrices by boldface upper
case letters, such as $\mathbf{X}$. Table~\ref{tab:notation} lists the symbols and their descriptions that are used throughout this paper.

More formally, a DCG $G$ consists of two types of nodes, the document nodes $\{d_i\}$ and the biomedical concept nodes $\{c_j\}$. A document $d_i$ is composed of a sequence of $k$ words $(x_1^{(i)}, x_2^{(i)}, ..., x_k^{(i)} )$ and contains $n$ medical concepts $(c_1^{(i)}, c_2^{(i)}, ..., c_n^{(i)})$. A specific biomedical concept $c_j$ can be composed of a single word or multiple words, which corresponds to a medical named entity or a key topic in medical domain. We link a concept node $c_j$ to a document node $d_i$ if this document contains the concept $c_j$. We link two concept nodes $c_u$ and $c_v$ if there exists any relations between them in external knowledge bases. Given the document-concept graph $G$, our objective is to learn the representation for each document node.

\subsection{Document-Concept Graph Construction}
\begin{figure}[tb]
	\centering
	\includegraphics[width=0.45\textwidth]{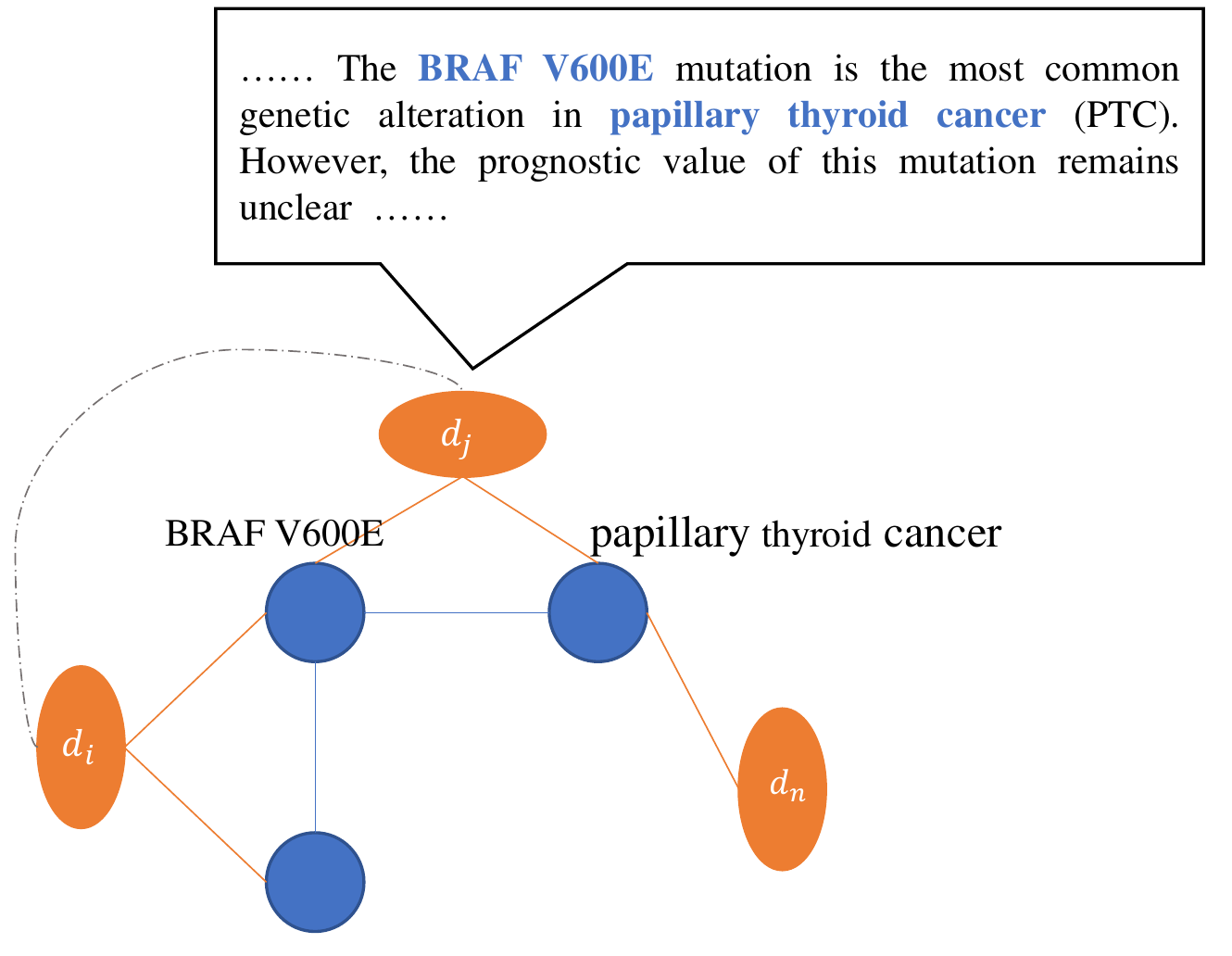}
	\caption{An example of the document-concept graph.}
	\label{fig:DCG}
\end{figure}

Figure~\ref{fig:DCG} illustrates an example of DCG. The construction of a DCG consists of four steps: 1) tokenization; 2) word embedding; 3) medical concept extraction; 4) edge construction. 

\underline{\emph{Tokenization}}. Given a biomedical article, tokenization is necessary to convert the texts into space-separated sequences of words (words may include punctuation or be followed by punctuation). In our work, the Stanford CoreNLP \cite{manning2014ACL} is leveraged to conduct tokenization on documents. 

\underline{\emph{Word Embedding}}. After tokenization, unsupervised word representation learning models like Glove \cite{pennington2014glove} and Word2Vec \cite{mikolov2013distributed} are implemented on all documents to get semantic representation for each word which would be useful in later parts. 

\underline{\emph{Medical Concept Extraction}}. Since each document might contain multiple medical concepts corresponding to different types of biomedical named entities, such as genes, diseases, chemicals, mutations, etc., which can be detected through medical named recognition and normalization methods and tools \cite{wei2013pubtator,zhao2018neural}. 

\underline{\emph{Edge Construction}}. Given all detected medical concepts in documents, we can construct the edges in a DCG. It is straightforward to link the document and concept nodes by just checking whether the concept is included in the document as the same way in \cite{zhao2017constructing}. For the edges linking concept nodes, we extract them from external knowledge bases as in \cite{zhu2018drug}. 

\subsection{Document Representation Learning}
\begin{figure}[tb]
	\centering
	\includegraphics[width=0.49\textwidth]{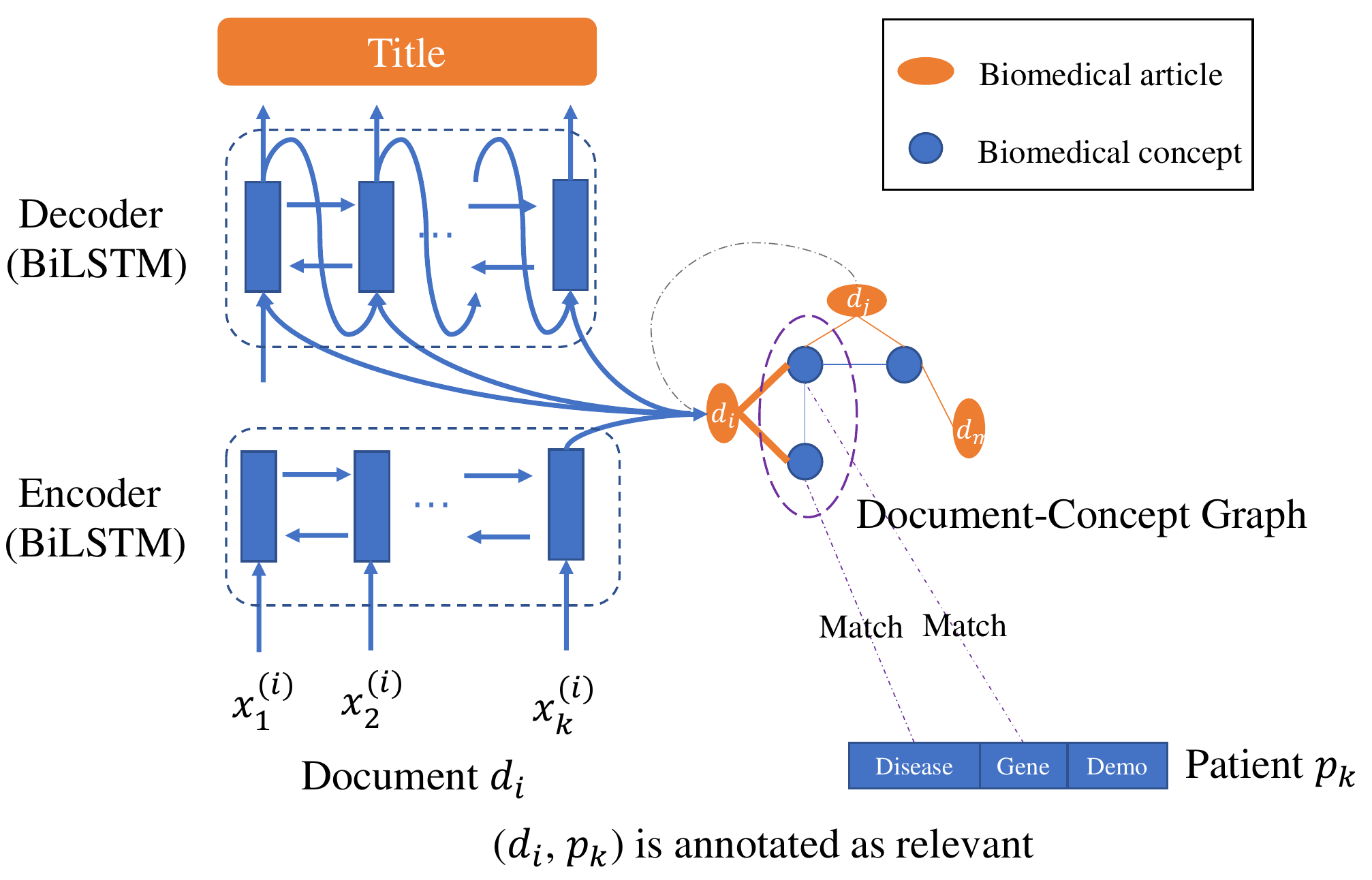}
	\caption{Graph-augmented RNN for document representation learning.}
	\label{fig:RNN-GCN}
\end{figure}

Given the DCG $G$, our next step is to learn effective representations for documents, which is a so crucial factor for the performance of our system. In particular, we propose to learn such representation through a graph (DCG) augmented recurrent neural network (RNN) model, which is shown in Figure~\ref{fig:RNN-GCN}. This type of Graph-LSTM framework has been applied in information extraction tasks \cite{qian2018graphie,song2018n}. 
Our model has three components:
\begin{itemize}
	\item \textit{Encoder}, which generates context-aware hidden representations for each biomedical document with a recurrent neural network;
	\item \textit{Graph Module}, which is essentially the DCG capturing the relationships among the documents and biomedical concepts as we introduced in the last subsection. 
	\item \textit{Decoder}, which exploits the contextual information generated by the encoder and the graph to predict key topics of the documents. Our goal is to predict the title for the abstract of each biomedical article.
\end{itemize}

\subsubsection{Encoder}
We first use an encoder to generate document representations. Given a document $d_i$ with word sequence $(x_1^{(i)}, x_2^{(i)}, ..., x_k^{(i)})$ of length $k$, each word $x_k^{(i)}$ is represented by a vector $\mathbf{x_k^{(i)}}$, which is obtained from some pre-trained word embedding.  We encode the document with a recurrent neural network defined as
\begin{equation}
\mathbf{h}_{1:k}^{(i)} = \text{RNN}(\mathbf{x}_{1:k}^{(i)};\mathbf{0};\Theta_{\text{enc}})
\end{equation}
where $\mathbf{x}_{1:k}^{(i)}$ represents the input word sequence $(x_1^{(i)}, x_2^{(i)}, ..., x_k^{(i)})$, $\mathbf{h}_{1:k}^{(i)}$ denotes the hidden states $[\mathbf{h}_1^{(i)}, \mathbf{h}_2^{(i)}, ..., \mathbf{h}_k^{(i)}]$. $\mathbf{0}$ denotes an all-zero vector. $\Theta_{\text{enc}}$ denotes parameters of the encoder. We implement the RNN as a bi-directional LSTM \cite{hochreiter1997long} to encode each document. 

We construct the textual representation for document $d_i$ by averaging the hidden states of its words, i.e. $\text{Enc}(d_i)=\frac{1}{k}(\sum_{t=1}^{k}\mathbf{h}_t^{(i)})$. These learned representations are then fed into the DCG as the initial representation of the document nodes.  The final document representation will be the combination of the textual representation and the structural representation induced from the DCG. In the following, we introduce how such structural representations are learned.

\subsubsection{The Graph Module}
The graph module is designed to get the structural representations for document nodes via message passing in the DCG $G$.  
Specifically, we adopt a restricted graph convolutional network (GCN) to achieve this goal.

Given the DCG $G = (V, E)$, where for a document node $v_{d_i}$ (representing document $d_i$), it has the encoding $\text{Enc}(d_i)$ capturing its textual information. For a concept node $v_{c_j}$ (representing biomedical concept $c_j$), it has a pre-trained word embedding representation. The graph module enhances such representation with the graph structure in DCG. More concretely, it operates on local neighborhoods in the graph to integrate medical concepts information to document nodes.
This means that we want the information flow from concept nodes to document nodes and the information flow between concept nodes, but we do not want the concept node to be influenced by the information from document nodes since we assume the semantic of a medical concept is much more concrete and stable than the document. 

Precisely, at each layer of the GCN, each document node gets information from its neighboring concept nodes, i.e., 
\begin{equation}
\mathbf{g}_{d_i}^{(l+1)} = \delta\left (\sum_{m\in \mathcal{M}_i}\alpha f_{m}(\mathbf{g}_{d_i}^{(l)}, \mathbf{g}_{c_j}^{(l)})\right ) 
\end{equation}
where $\mathbf{g}_{d_i}^{(l)} \in \mathbb{R}^{d^{(l)}}$ is the hidden state of document node $v_{d_i}$ in the $l$-th layer of the neural network, with $d^{(l)}$ being the dimensionality of this layer's representation. Incoming messages from $f_m$ are accumulated and passed through an element-wise activation function $\delta(.)$, such as ReLU. $\mathcal{M}_i$ denotes the set of incoming messages for document node ${v}_{d_i}$. $f_m(.,.)$ is typically chosen to be a (message-specific) neural network-like function or simply a linear transformation $f_m(\mathbf{g}_{d_i}^{(l)}, \mathbf{g}_{c_j}^{(l)})= \mathbf{W}_d\mathbf{g}_{c_j}$ with a weight matrix $\mathbf{W}_d$. $\alpha$ is a reward factor. The neighboring biomedical concepts of an article may differ in their importance. Therefore, it is reasonable to assign different weights to those medical concepts in sending message to their central document node. For those neighboring concept nodes which do not match any related queries, we set $\alpha = 1$. For those neighboring concept nodes which match related queries, we set $\alpha>1$.

At each layer of the GCN, each concept node get information from its neighboring concept nodes, i.e., 
\begin{equation}
\mathbf{g}_{c_i}^{(l+1)} = \delta\left (\sum_{n\in \mathcal{N}_i}f_{n}(\mathbf{g}_{c_i}^{(l)}, \mathbf{g}_{c_j}^{(l)})\right ) 
\end{equation}
where $\mathbf{g}_{c_i}^{(l)} \in \mathbb{R}^{d^{(l)}}$ is the hidden state of concept node $v_{c_i}$ in the $l$-th layer of the neural network, with $d^{(l)}$ being the dimensionality of this layer's representation. Incoming messages of the form $f_n$ are accumulated and passed through an element-wise activation function $\delta(.)$, such as ReLU. $\mathcal{N}_i$ denotes the set of incoming messages for concept node ${v}_{c_i}$. $f_n(.,.)$ is typically chosen to be a (message-specific) neural network-like function or simply a linear transformation $f_n(\mathbf{g}_{c_i}^{(l)}, \mathbf{g}_{c_j}^{(l)})= \mathbf{W}_c\mathbf{g}_{c_j}$ with a weight matrix $\mathbf{W}_c$.

In each layer of the GCN, information is only propagated through neighboring nodes that are directly connected. Thus we can stack more GCN layers to get a larger node receptive field, i.e. each node can get information from more distant neighbors through transitivity on the graph. After $L$ layers, for each document node $v_{d_i}$ we obtain its structural representation $\text{GCN}(d_i) = \mathbf{g}_{d_i}^{(L)}$. Such representation will be combined with its corresponding textual representation $\text{Enc}(d_i)$ to form the final representation of each document, i.e.
\begin{equation}
\mathbf{v}_{d_i} = f_l(\text{Enc}(d_i)\circ \text{GCN}(d_i))
\end{equation}
where $f_l$ is a linear transformation and $\circ$ is the concatenation operator. In this way the representation captured both the local textual information (induced from the encoder) and the global topic information (learned from the GCN) of each document. 

\subsubsection{Decoder}
To make sure document representations can encode relevant local and non-local concepts, we design a decoder to see if key relevant information can be reproduced. The decoder takes as input the representation of each document $\mathbf{v}_{d_i}$ and generates a title for the document. 
In particular, given the document representation $\mathbf{v}_{d_i}$, we apply a recurrent neural network to generate a word or a medical concept once a time from a controlled vocabulary including words and medical entities. The decoder is defined as 
\begin{equation}
w_{1:L}^{(i)} = \text{RNN}(\mathbf{h}_{1:L}^{(i)},\mathbf{v}_{d_i}, \Theta_{\text{dec}}) 
\end{equation}
$w_{1:L}^{(i)}$ represents the generated word/entity sequence, $L$ is the set max length of generated sequence. The process would stop early when `EOS' (end-of-sequence) is generated, $\mathbf{h}_{1:L}^{(i)}=[\mathbf{h}_1^{(i)}, \mathbf{h}_2^{(i)}, ..., \mathbf{h}_L^{(i)}]$ denotes the hidden states. $\mathbf{v}_{d_i}$ serves as the initialization for each state, $\Theta_{\text{dec}}$ denotes parameters of the decoder. We implement the RNN as a bi-directional LSTM to generate word/entity sequence. 

In this way, we can construct the loss to optimize over the model parameters using the sequence cross-entropy loss shown as follows
\begin{equation}
\mathcal{L}_{\text{graph}} = -\frac{1}{IL}\sum_{i = 1}^{I}\sum_{l=1}^{L}\log p(w_l|d_i,w_{<l};\Theta)
\end{equation}
where $I$ is the number of documents, $L$ is the max length of generated sequence of word/entity, $\Theta$ represents all parameters involved in encoder, graph module, and decoder.

\section{Learning to rank documents}
In this section, we propose a learning-to-rank model to identify relevant biomedical articles for a given query (e.g., patient with genetic information in terms of treatment, prevention, and prognosis of the disease) with the document representation vector learned. First, we discuss how to effectively represent a query.

\subsection{Representation learning for query}
\begin{figure}[tb]
	\centering
	\includegraphics[width=0.4\textwidth]{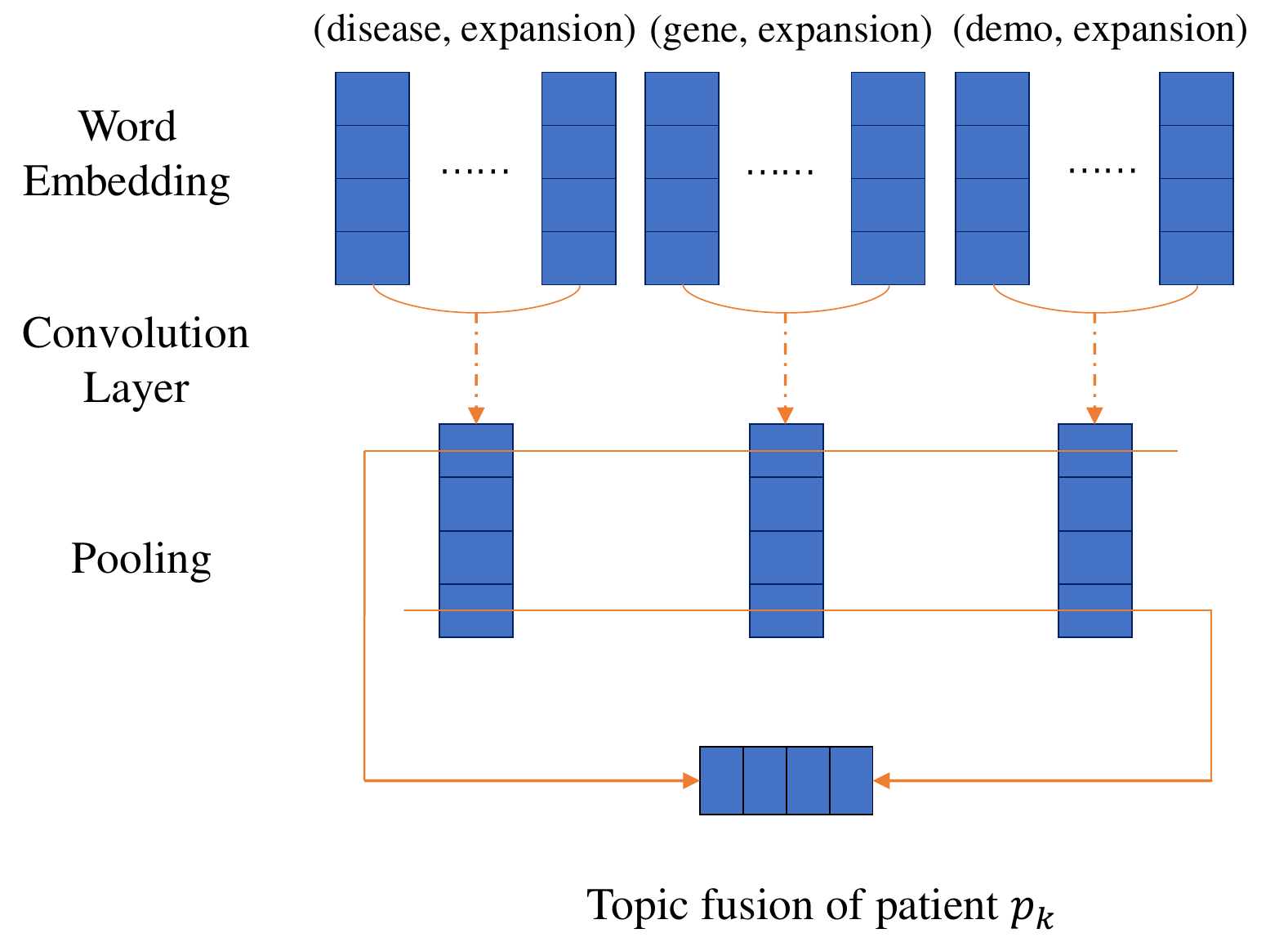}
	\caption{Query representation learning model based on CNN.}
	\label{fig:CNN}
\end{figure}

Each query contains information including the patient's disease name, the relevant genetic variants (which genes), basic demographic information (age, sex), etc.
It is very common that the same disease and variant have different expressions. For example `leucocythemia' and `leukemia' are the same disease but expressed in different ways. Disease names are frequently expressed using abbreviations in literature. Different institutions may name the same genetic variants in different ways. All these aspects make it challenging to match the query to relevant biomedical articles. Therefore, it is necessary to expand queries to incorporate these different expressions.

To get the flexible representation of the query, we propose to utilize a variety of knowledge bases in Table~\ref{tab:kb} to carefully expand gene, variant, and disease terms.
\begin{table}[htb]
	\caption{Knowledge bases used for query expansion}
	\label{tab:kb}
	\begin{tabularx}{0.48\textwidth}{cX}
		\toprule
		Expansion &Knowledge bases\\
		\midrule
		gene name expansion & NCBI GeneDB, HGNC, COSMIC, Entrez Gene Library, NCBI Homo Sapiens, PMDG\\
		disease expansion & NCI thesaurus, MeSH hierarchy, SNOMED/Lexigram, SNOMED CT\\
		variant expansion  & COSMIC\\
		\bottomrule
	\end{tabularx}
\end{table}

Since the query is given as structured patient information including disease, variant, and demographic, it is crucial to combine them together and represent as a vector. 
We exploit convolutional neural network (CNN) to convolve structured items and obtain the vector representation of the query. Note that this part shares the same pre-trained word embedding vocabulary as is used in the document representation learning part. The CNN model is shown in Figure~\ref{fig:CNN}.

For the query of MeSH terms, we expand each term via knowledge bases in Table~\ref{tab:kb} and apply CNN to convolve each Mesh term and its expansion so as to share the same CNN model with clinical topics.

\subsection{Learning to rank biomedical articles}
\label{sec:method}
For each $(p_k,D)$ pair, where $D$ is a ranking list $(d^1,d^2,d^3, ......)$ of biomedical articles with the superscript indicating the ranking order of the document and $p_k$ is the $k$-th query, we define a relevant scoring function $f(p_k, d^i)$ as
\begin{equation}
f(p_k,d_i) = \left\|\mathbf{W}\times\mathbf{v}_{p_k} - \mathbf{v}_{d_i} \right\|_2
\end{equation}
where we use an $\ell_1$ norm in the latent space, but other metrics could be used as well, $\mathbf{v}_{d_i}$ is the vector representation obtained from document representation learning and $\mathbf{v}_{p_k}$ is the vector representation obtained from patient representation learning, $\mathbf{W}$ is linear transformation matrix. The scoring function evaluates the relevance between $p_k$ and $d^i$, smaller value indicates higher relevance between $p_k$ and $d^i$ . Therefore, we can generate the following pairs of partial order from the sample $(p_k, D)$:
\begin{equation}
\begin{split}
&f(p_k, d^1 )<f(p_k, d^2 );\\
&f(p_k, d^1 )<f(p_k, d^3 );\\
&f(p_k, d^2 )<f(p_k, d^3 );\\
&......\\
\end{split}
\end{equation}
With these pairs, we can define the loss of pairwise learning to rank as follows
\begin{equation}
\mathcal{L}_{\text{rank}}= \frac{1}{KIJ}\sum_{k=1}^K \sum_{i=1}^I\sum_{j>i}^J max(0, f(p_k,d_{p_k}^{i})- f(p_k,d_{p_k}^j))   
\end{equation}
This is a margin loss to make sure the score of $d_{p_k}^{i}$ is smaller than $d_{p_k}^j$ when $j>i$, where $d_{p_k}^{i}$ denotes the document is ranked as $i$-th for the query $p_k$.  Our object is to minimize total loss on all training data. For a new query, the biomedical articles would be ranked according to  the value of scoring function. The ranking model is trained to give more relevant documents a smaller score by tuning its parameters to minimize the pairwise maximum margin loss.

Therefore, the total loss of the entire ranking framework GRAPHENE is defined as
\begin{equation}
\mathcal{L} = \beta\mathcal{L}_{\text{graph}} + \gamma\mathcal{L}_{\text{rank}}   
\end{equation}
where $\beta = 0.5$ and $\gamma =0.5$ are hyper-parameters. 
The Adagrad stochastic gradient descent method is used to train the model with mini-batch mode. 

\section{Experiments}
This section describes the details of our empirical study on evaluating our proposed framework, including the ranking benchmark dataset, the experimental settings, and results.

\subsection{Data set}
We use the dataset from TREC Precision Medicine track (\url{http://www.trec-cds.org/}, 2017 and 2018) clinical scenarios and medical articles for empirical evaluation. We also exploited the MeSH\footnote{MeSH (Medical Subject Headings, \url{https://www.nlm.nih.gov/mesh/meshhome.html}) is the National Library of Medicine's controlled vocabulary thesaurus. Each biomedical article is associated with a set of MeSH terms describing the content of the citation.} terms of each biomedical article in the first dataset as query and take the corresponding article to be the unique relevant document to be retrieved.
The TREC dataset comprises 80 clinical scenarios (called topics) with structured patient scenarios including information related to disease, genetic variants, demographics, and other relevant factors. The dataset was curated by precision oncologists from MD Anderson. Therefore, it included relevant information about cancer patients such that participant systems can retrieve pertinent biomedical articles. The biomedical article corpus is composed of approximately 26.8 million MEDLINE abstracts with titles and associated MeSH terms. It is supplemented with two additional sets of abstracts: (i) 37,007 abstracts from recent proceedings of the American Society of Clinical Oncology (ASCO), and (ii) 33,018 abstracts from recent proceedings of the American Association for Cancer Research (AACR). These additional datasets were added to increase the set of potentially relevant treatment information. One reason for this is because the fact that the latest research is often presented at conferences such as ASCO and AACR prior to submission to journals (thus these proceedings may represent a more up-to-date snapshot of scientific knowledge than MEDLINE).

For each clinical topic, there is a list of relevant biomedical articles with human-labeled relevance scores. Therefore, it is very convenient to generate a ranking list of articles for each topic. We take 50 clinical topics of TREC PM track 2018 as the training set and the left 30 clinical topics of TREC PM track 2017 as the testing set.
Due to the limited number of clinical topics for training a deep neural IR model, we use the MeSH terms of each biomedical article as the corresponding query to pre-train deep neural models. In the pre-training stage, for each unique $\langle$MeSH term, article$\rangle$ pair, we randomly sample a different article to generate a controlled pair $\langle$MeSH terms, unrelated article$\rangle$.

For each $\langle$MeSH term, article$\rangle$ pair, we can take the MeSH terms as query and search the corresponding article. 
Since the Mesh terms and article are one-one co-related. We can take the MeSH terms as another query set to conduct biomedical literature retrieval task. We split all pairs of $\langle$MeSH, article$\rangle$ and corresponding $\langle$MeSH term, unrelated article$\rangle$ into a training/validation/test set randomly, with the ratio of 8:1:1. The first part is for the training model, the second for hyper-parameter tuning, and the third for evaluation.
Since there is only one matched article for a particular MeSH term query in most cases.  Therefore, it is more reasonable to use {Prec.1 and MRR instead of Prec.10 and NDCG.20} (which are explained in detail below) as the metrics to evaluate the performance on the query set of MeSH terms.

\subsection{Evaluation Metrics and Settings}
We introduce a set of metrics for evaluating the retrieval performance of the algorithms. All metrics have values in the range of [0, 1], with higher values for better rankings. When generating the ranked list, for documents with the same relevance scores, they will be ranked according to their IDs.

\subsubsection{NDCG}
Discounted Cumulative Gain (DCG) \cite{jarvelin2002cumulated} is a relevance and rank correlation metric that penalizes placement of relevant documents at lower ranks. The traditional formula of DCG accumulated at a particular rank position $n$ is defined as:
\begin{equation}
DCG(n) = \sum_{i=1}^n\frac{rel_i}{\log_2(i+1)}
\end{equation}
Where $rel_i$ is the graded relevance of the result at position $i$.  Normalized Discounted Cumulative Gain (NDCG) then measures the relative
DCG of a ranking compared to the best possible ranking for that data: $NDCG(n) = DCG(n)/IDCG(n)$, where $IDCG(n)$ is the $DCG(n)$
for the ideal ranking. When there are multiple queries, NDCG refers to the mean value across queries. We use the scaled relevance levels, and quote ``NDCG.20" metrics for $n = 20$.

\subsubsection{Precision at Rank, MRR and MAP}
{Average Precision measures, for a single query, is the weighted sum of the precision observed in a ranked list up to each specific rank weighted by the actual relevance score of the corresponding document, averaged over the number of relevant documents for that query}. It is thus a ranking measure that factors out the size of the ranked list and the number of relevant documents, without any rank-based penalization or discounting. The Mean Average Precision (MAP) is the mean of the Average Precision across queries in our test dataset. The Precision at rank n metrics (``Prec.n") is the retrieval precision at rank n. The mean reciprocal rank (MRR) is the multiplicative inverse of the rank of the first correct answer.

\subsubsection{Implementation Details}
In the experiments, we evaluate the performance of GRAPHENE in two query sets for retrieving biomedical articles, including (1) clinical topics, and (2) MeSH terms. Our RNN network is a 3-layer BiLSTM with pre-trained word embeddings. Hyperparameters of our 2-layer GCN are set by the same values reported in Kipf and Welling \cite{kipf2016semi}. We follow the training procedure outlined in Section~\ref{sec:method} with the word embeddings setup in Section~\ref{sec: embeddings}. The best setting of the reward factor $\alpha$ for those matched concept nodes is $1.6$. We use a dropout rate of 0.5 and train the model with Adagrad SGD with the initial learning rate of 0.001 and momentum of 0.9 for 50 epochs. For baseline deep neural models, we use the same setting reported in their corresponding papers.

\subsection{Baselines}
\begin{table*}[h]
	\small
	\caption{Comparison of different retrieval models over the query sets MeSH terms and clinical topics. The value with * is infNDCG value.}\label{tab: base}
	\centering
	\begin{tabular}{cccccccc}
		\hline
		\multirow{2}{*}{Model Type}&\multirow{2}{*}{Model Name} &\multicolumn{3}{c}{MeSH terms} & \multicolumn{3}{c}{Clinical topics}\\
		\cmidrule(lr){3-5} \cmidrule(lr){6-8} 
		\multirow{2}{*}{}&\multirow{2}{*}{}&  Prec.1 & MRR & MAP & NDCG.20 & MAP & Prec.10\\
		\hline
		\multirow{2}{*}{Traditional Retrieval Models}& BM25 &0.0927&0.2364&0.2359&0.2832&0.0908& 0.2967\\
		\multirow{2}{*}{}&DLH13 &0.1374&0.2873&0.2856&0.4726& 0.1280 & 0.4800\\
		\hline
		\multirow{2}{*}{Deep Neural IR Models}&DeepMatch &0.2111&0.3604&0.3589 &0.6081&0.4375&0.6132\\
		\multirow{2}{*}{}&Delta&0.3136&0.3897&0.3842 &0.6796&0.4856&0.6831\\
		
		\hline
		\multirow{2}{*}{TREC PM track}&UD\_GU\_BioTM &-&-&- &0.4135*&- &0.6400\\
		\multirow{2}{*}{}&UTD HLTRI &-&-&- & 0.4593* &- & 0.6172\\
		
		\hline
		\multirow{3}{*}{Our Approach}&GRAPHENE(-graph)&0.2745&0.3712&0.3674 &0.6138&0.4693&0.6342\\
		\multirow{3}{*}{}&GRAPHENE(-reward)&0.3354&0.4269&0.4243&0.6736 & 0.5732& 0.6837\\
		\multirow{3}{*}{}&GRAPHENE&0.3356&0.4296&0.4276&0.6907& 0.5967& 0.7024\\
		\hline
	\end{tabular}	
\end{table*}

We compared the performance of GRAPHENE with the following IR models as baselines.

\textbf{BM25} \cite{robertson1995okapi} is a bag-of-words retrieval model that ranks a set of documents based on the appearance of the query terms in each document, regardless of the intra-relationship between the query terms within a document (e.g., their relative proximity).

\textbf{DLH13 (DFR)} \cite{amati2006frequentist}  is a probabilistic retrieval model based on vector matching. However, they replace the term frequencies in the bag-of-words model with the probabilities inferred from a fitted Poisson model. 

\textbf{DeepMatch} \cite{lu2013deep} is a deep learning architecture aiming at capturing the complicated matching relations between two objects from heterogeneous domains more effectively. DeepMatch is an interaction-focused model. It directly modeled the object-object interactions with a CNN-based architecture. In particular, it convolves on object-object interaction matrix and predicts if two objects are related. 

\textbf{Delta} \cite{mohan2018fast} constructs a ``modified" document matrix first by replacing the words in the documents by the closest words in the query. Convolutions will be performed on this matrix to obtain a final relevance score. 


\subsection{Pre-trained word embeddings}
\label{sec: embeddings}
We initialized the word embedding matrix with three types of pre-trained word embeddings respectively. 
The first is Word2Vec 100 dimensional embeddings trained on all 27.5 million MEDLINE biomedical articles in our data. The second is GloVe 100 dimensional embeddings trained on the same entire 27.5 million MEDLINE biomedical articles. The third is the randomly initialized 100 dimensional embeddings which are uniformly sampled from range $[-\sqrt{\frac{3}{dim}},+\sqrt{\frac{3}{dim}}]$, where $dim$ is the dimension of embeddings~\cite{He:2015:DDR:2919332.2919814}.

\subsection{Results}
This section presents the performance results of different retrieval models over the two benchmark query sets. The overall summary of the results is displayed in Table~\ref{tab: base}.

From the table we can observe that 
\begin{itemize}
	\item The deep neural IR models, Delta and DeepMatch, perform significantly better than the traditional retrieval models. This verifies the advantage of these deep neural models for retrieving relevant documents compared to retrieval models based on simple matching with one-hot representations.
	
	\item The TREC PM track models are the top two results reported at TREC precision track 2017. Since these two models utilize many external knowledge bases, rules, and ensemble strategies, these top two TREC PM track models produce very good results.
	
	\item Our GRAPHENE system achieves the best performance.
\end{itemize}

In order to test the necessity of the different components in GRAPHENE, we also implemented two variants. GRAPHENE(-graph) excludes the graph module in document representation learning part. GRAPHENE(-reward) ignores the reward part when the passing message from concept nodes to document nodes by assigning equal weights on all messages. From the results we can see that both variants perform worse than the original GRAPHENE, which indicates that all components in GRAPHENE are important and necessary.


Moreover, we can also observe that GRAPHENE performs the best on the query set of MeSH terms, which is a one-one matching problem, which means that given a set of MeSH terms only one biomedical article should be matched in most cases. This also demonstrates the potential of our model in precise searching problems such as question answering, email search, citation recommendation, etc.

\begin{table}[h]
	\small
	\caption{Performance with different choices of pre-trained word embeddings on deep neural IR models}
	\label{tab:embeddings}
	\small
	\begin{tabular}{llcc}
		\toprule
		\multirow{2}{*}{Word Embedding} & \multirow{2}{*}{Model}&\multicolumn{2}{c}{MAP}\\
		\cline{3-4}
		\multirow{2}{*}{} & \multirow{2}{*}{}&MeSH terms & Clinical topics\\
		\midrule
		\multirow{3}{*}{Random} &DeepMatch   &0.1826 & 0.2723\\
		\multirow{3}{*}{}&Delta              &0.3276 & 0.4054\\
		\multirow{3}{*}{}&GRAPHENE           &0.3629 & 0.4362\\
		\hline
		\multirow{3}{*}{Word2Vec} &DeepMatch  &0.33326 & 0.4155\\
		\multirow{3}{*}{}&Delta               &0.3521 & 0.4645\\
		\multirow{4}{*}{}&GRAPHENE            &0.4131 & 0.5903\\
		\hline
		\multirow{3}{*}{Glove} &DeepMatch     &0.3589&0.4375\\
		\multirow{3}{*}{}&Delta               &0.3842&0.4856\\
		\multirow{3}{*}{}&GRAPHENE            & 0.4276 & 0.5967\\
		\bottomrule
	\end{tabular}
\end{table}
We have also tested the effects of initialization with different strategies for pre-training the word embeddings described in Section~\ref{sec: embeddings}. The results are shown in Table~\ref{tab:embeddings}. From the table we can observe that 
\begin{itemize}
	\item Models using pre-trained word embeddings achieve a significant improvement as opposed to the ones using random embeddings. 
	\item Models using GloVe embeddings outperforms using Word2Vec consistently for different neural models in different data sets. 
	\item DeepMatch and Delta rely more heavily on pre-trained word embeddings compared to our proposed GRAPHENE.
\end{itemize}
The reason why pre-trained word embeddings affect less on our GRAPHENE is probably due to the encoder-decoder part of GRAPHENE to generate title can {provide training of more customized word embeddings on large-scale abstract-to-title samples}.
\begin{figure}[tb]
	\centering
	\includegraphics[width=0.45\textwidth]{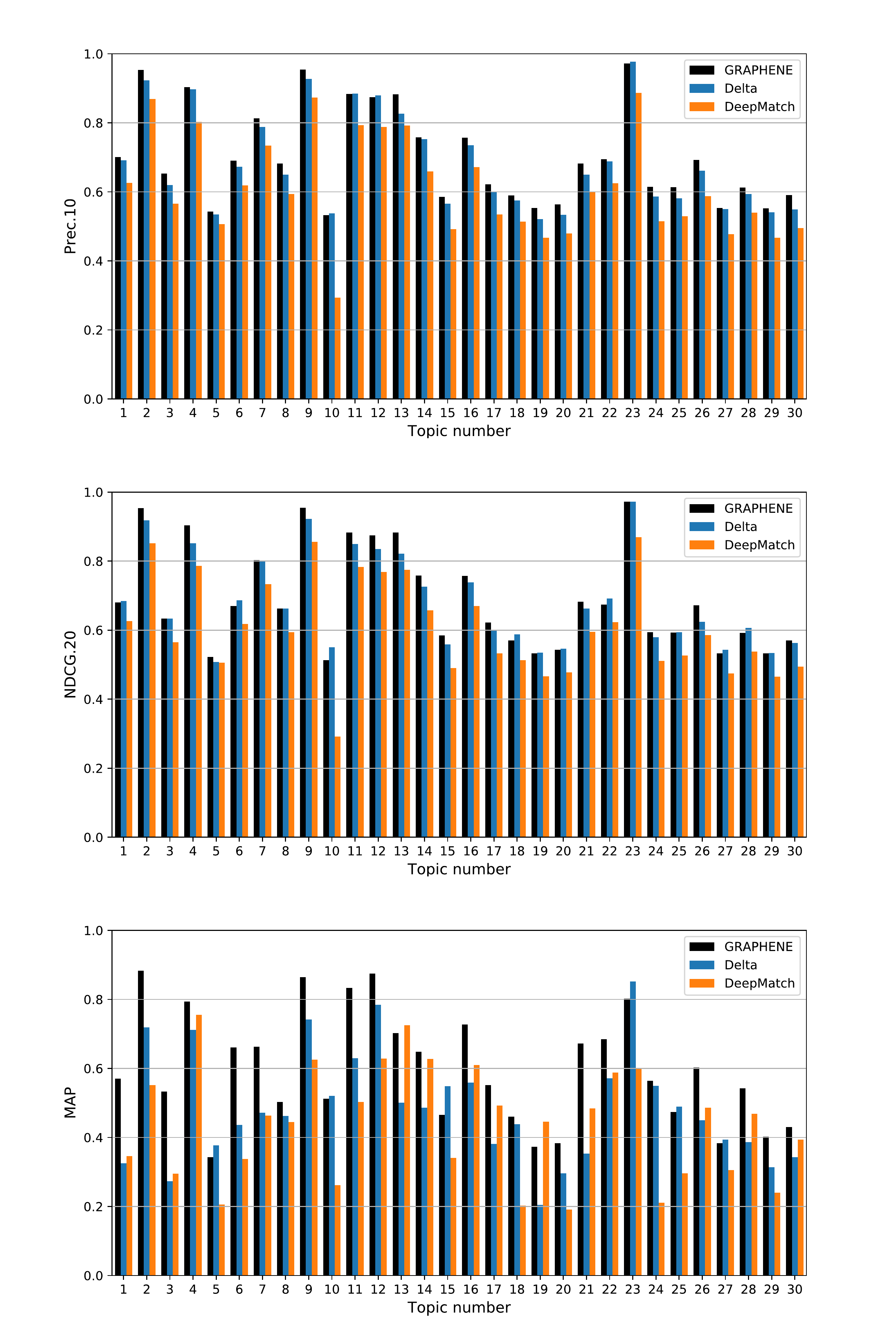}
	\caption{Prec.10 scores, NDCG.20 scores and MAP scores for each clinical topic.}
	\label{fig:NDCG}
\end{figure}

Moreover, we also investigated the importance of pre-training via the query set of MeSH terms for biomedical literature retrieval. We have performed experiments without pre-training on `MeSH terms' and directly trained the model on 50 clinical topics. According to the results in Table~\ref{tab:MeSH}, deep neural IR models without pre-training using MeSH terms query set drop sharply on the performance of biomedical literature retrieval. Compared to DeepMatch, the performance of Delta and our proposed GRAPHENE have a greater dependency on the pre-training with large-scale data. 
\begin{table}[h]
	\small
	\caption{Performance without pre-train via MeSH terms on deep neural IR models}
	\label{tab:MeSH}
	\begin{tabular}{cccc}
		\toprule
		Model&NDCG.20 & MAP & Prec.10\\
		\midrule
		DeepMatch     &0.1767&0.1265&0.1923\\
		Delta           &0.1134&0.0843&0.1243\\
		GRAPHENE       & 0.1221 & 0.0752 & 0.1385\\
		\bottomrule
	\end{tabular}
\end{table}

\begin{figure}[tb]
	\centering
	\includegraphics[width=0.45\textwidth]{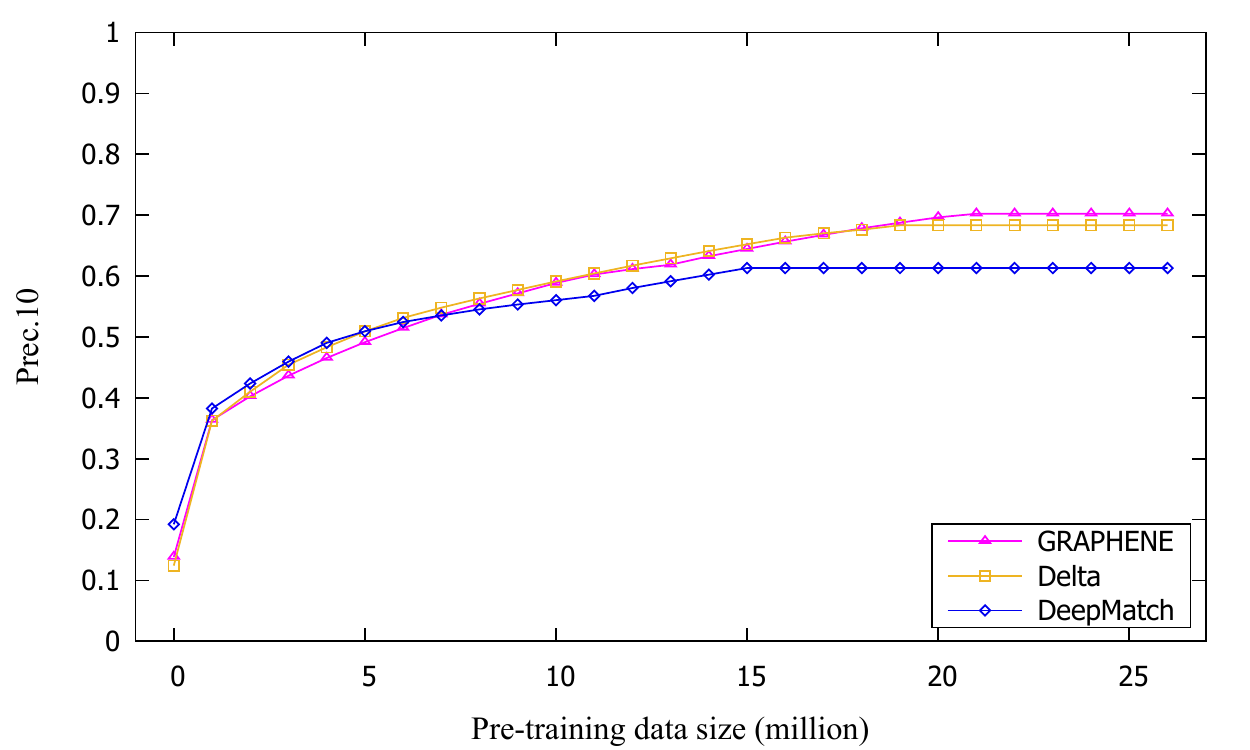}
	\caption{Comparison of deep neural IR models by incrementally augmenting pre-training query set of MeSH terms.}
	\label{fig:pretrain}
\end{figure}
Figure~\ref{fig:NDCG} shows the overall scores of our model for biomedical literature retrieval across all 30 clinical topics as compared to Delta and DeepMatch. From the figures we can see that GRAPHENE performs consistently better than the baseline models for nearly all the topics across all evaluation measures on the query set of clinical topics. 
Delta outperforms GRAPHENE in 4 clinical topics and DeepMatch outperforms Delta in 3 clinical topics. Here, it is worth to mention that the first and second  sub-figures in Figure~\ref{fig:NDCG} show the evaluation on top n documents and the third sub-figure in Figure~\ref{fig:NDCG} shows the evaluation on all related documents, and the evaluation on top n results is much more stable than the evaluation on all related documents.

The last question we investigated is whether the size of pre-training query set of MeSH terms would affect the final performance. As shown in Figure~\ref{fig:pretrain}, we check the performance of deep neural IR models on biomedical literature retrieval of clinical topics by incrementally augmenting pre-training data. The results indicate clearly that the size of pre-training data plays a key role in all three deep neural IR models, and it is more critical to GRAPHENE and Delta, both of which are  representation-focused models rely more heavily on large scale of pre-training data. This is consistent with the results of previous work \cite{xiong2017end}.

\begin{table*}[htb]
	\centering
	\small
	\caption{Examples of matched documents ranked as top 1 by different deep neural IR models. Bold blue text is the medical concepts which appear in the query, italic blue text is the related medical concept.}
	\begin{tabularx}{\textwidth}{|l|l|X|}
		\hline
		Query & Model & Document\\
		\hline
		\multirow{3}{*}{}  &GRAPHENE   &We have previously shown that \textit{\textcolor{blue}{arsenic trioxide}} blocks proliferation and induces apoptosis in human \textbf{\textcolor{blue}{pancreatic cancer}} cells at low, ...... expression of CDK2, CDK4, \textbf{\textcolor{blue}{CDK6}}, and cyclin E were not affected ...... In summary, \textit{\textcolor{blue}{arsenic trioxide}} induced apoptosis in \textbf{\textcolor{blue}{ pancreatic cancer}} cells through activating the caspase cascade via the mitochondrial pathway ...... This old drug may be valuable for treatment of pancreatic cancer.\\
		\cline{2-3}
		\multirow{3}{*}{\tabincell{l}{Pancreatic cancer;\\ CDK6 Amplification;\\  48-year-old male}}  & Delta & ...... An association has been reported between p16 mutations and \textbf{\textcolor{blue}{ pancreatic cancer}} ...... The second most frequent cancer was \textbf{\textcolor{blue}{ pancreatic cancer}} ...... of \textbf{\textcolor{blue}{ pancreatic cancer}} was  ......  The estimated cumulative risk of developing \textbf{\textcolor{blue}{ pancreatic cancer}} in putative mutation ...... no cases of \textbf{\textcolor{blue}{ pancreatic cancer}} occurred. p16 mutation carriers have a considerable risk of developing \textbf{\textcolor{blue}{ pancreatic cancer}} ...... \\
		\cline{2-3}
		\multirow{3}{*}{}  & DeepMatch & ...... An association has been reported between p16 mutations and \textbf{\textcolor{blue}{ pancreatic cancer}} ...... The second most frequent cancer was \textbf{\textcolor{blue}{pancreatic cancer}} ...... of \textbf{\textcolor{blue}{ pancreatic cancer}} was  ......  The estimated cumulative risk of developing \textbf{\textcolor{blue}{ pancreatic cancer}} in putative mutation ...... no cases of \textbf{\textcolor{blue}{ pancreatic cancer}} occurred. p16 mutation carriers have a considerable risk of developing \textbf{\textcolor{blue}{ pancreatic cancer}} ...... \\
		\hline
	\end{tabularx}
	\label{tab:case}
\end{table*}

\textbf{Case Study:}
Last but not the least, we performed a case study to better understand the power of GRAPHENE. Table~\ref{tab:case} shows an example of retrieved relevant documents that are placed at rank 1 by GRAPHENE, DeepMatch and Delta, with respect to a specific query. We highlight the biomedical concepts which are contained in query in blue bold text and the relevant medical concepts with entity in query blue italic text. 
From the results we can see that in the document retrieved by GRAPHENE, there are matched disease name ``pancreatic cancer", matched gene variant ``CDK6" and the relevant medical concept ``arsenic trioxide" (treatment of the pancreatic cancer), while only the disease name ``pancreatic cancer" appears in the most relevant document identified by Delta and DeepMatch. This demonstrates the strong potential of GRAPHENE on identification of semantic similar items despite the different exact term expressions.

\section{Conclusion}
In this paper, we proposed a deep neural biomedical literature retrieval framework GRAPHENE which consists of graph-augmented document representation learning, query expansion, and representation learning, and learning to rank biomedical articles. The graph-augmented document representation learning is applied upon a document-concept graph which contains biomedical concept nodes and document nodes so that global biomedical concept co-occurrence can be explicitly modeled and graph convolution can be easily adapted. Query representation learning exploits a CNN-based model to convolve structured items of patients and output the vector representation for each query. A learning-to-rank algorithm is proposed to use partial order between biomedical articles with the given patient and learn the rank of relevant articles. Experimental results on TREC Precision Medicine track data provided compelling evidence to support that our model can effectively retrieve most relevant biomedical articles for a given query. The results also demonstrated that our GRAPHENE improves previous deep neural retrieval models which represent the document with textual information only, suggesting the necessity of encoding graph-based external knowledge bases into document representation. 

Since the great success of pre-trained BERT model in modeling long text for many NLP tasks recently \cite{devlin2018bert,beltagy2019scibert}, we want to integrate it into our present GRAPHENE to replace the RNN-based encoder-decoder module in our future work. Additionally, we also want to leverage advantages from both representation-focused models and interaction-focused models for our future work as they have respective advantages.

\section{Acknowledgments}
This work is supported by NSF 1716432 and 1750326.

%
\bibliographystyle{ACM-Reference-Format}
\bibliography{acmart}


\begin{thebibliography}{42}


\ifx \showCODEN    \undefined \def \showCODEN     #1{\unskip}     \fi
\ifx \showDOI      \undefined \def \showDOI       #1{#1}\fi
\ifx \showISBNx    \undefined \def \showISBNx     #1{\unskip}     \fi
\ifx \showISBNxiii \undefined \def \showISBNxiii  #1{\unskip}     \fi
\ifx \showISSN     \undefined \def \showISSN      #1{\unskip}     \fi
\ifx \showLCCN     \undefined \def \showLCCN      #1{\unskip}     \fi
\ifx \shownote     \undefined \def \shownote      #1{#1}          \fi
\ifx \showarticletitle \undefined \def \showarticletitle #1{#1}   \fi
\ifx \showURL      \undefined \def \showURL       {\relax}        \fi
\providecommand\bibfield[2]{#2}
\providecommand\bibinfo[2]{#2}
\providecommand\natexlab[1]{#1}
\providecommand\showeprint[2][]{arXiv:#2}

\bibitem[\protect\citeauthoryear{Amati}{Amati}{2006}]%
        {amati2006frequentist}
\bibfield{author}{\bibinfo{person}{Giambattista Amati}.}
  \bibinfo{year}{2006}\natexlab{}.
\newblock \showarticletitle{Frequentist and bayesian approach to information
  retrieval}. In \bibinfo{booktitle}{\emph{Proceedings of the ECIR}}.
  \bibinfo{pages}{13--24}.
\newblock


\bibitem[\protect\citeauthoryear{Beltagy, Cohan, and Lo}{Beltagy
  et~al\mbox{.}}{2019}]%
        {beltagy2019scibert}
\bibfield{author}{\bibinfo{person}{Iz Beltagy}, \bibinfo{person}{Arman Cohan},
  {and} \bibinfo{person}{Kyle Lo}.} \bibinfo{year}{2019}\natexlab{}.
\newblock \showarticletitle{SciBERT: Pretrained Contextualized Embeddings for
  Scientific Text}.
\newblock \bibinfo{journal}{\emph{arXiv preprint arXiv:1903.10676}}
  (\bibinfo{year}{2019}).
\newblock


\bibitem[\protect\citeauthoryear{Collins and Varmus}{Collins and
  Varmus}{2015}]%
        {collins2015new}
\bibfield{author}{\bibinfo{person}{Francis~S Collins} {and}
  \bibinfo{person}{Harold Varmus}.} \bibinfo{year}{2015}\natexlab{}.
\newblock \showarticletitle{A new initiative on precision medicine}.
\newblock \bibinfo{journal}{\emph{New England journal of medicine}}
  \bibinfo{volume}{372}, \bibinfo{number}{9} (\bibinfo{year}{2015}),
  \bibinfo{pages}{793--795}.
\newblock


\bibitem[\protect\citeauthoryear{Dai, Xiong, Callan, and Liu}{Dai
  et~al\mbox{.}}{2018}]%
        {Dai2018WSDM}
\bibfield{author}{\bibinfo{person}{Zhuyun Dai}, \bibinfo{person}{Chenyan
  Xiong}, \bibinfo{person}{Jamie Callan}, {and} \bibinfo{person}{Zhiyuan Liu}.}
  \bibinfo{year}{2018}\natexlab{}.
\newblock \showarticletitle{Convolutional Neural Networks for Soft-Matching
  N-Grams in Ad-hoc Search}. In \bibinfo{booktitle}{\emph{Proceedings of the
  WSDM}}. \bibinfo{pages}{126--134}.
\newblock


\bibitem[\protect\citeauthoryear{Devlin, Chang, Lee, and Toutanova}{Devlin
  et~al\mbox{.}}{2018}]%
        {devlin2018bert}
\bibfield{author}{\bibinfo{person}{Jacob Devlin}, \bibinfo{person}{Ming-Wei
  Chang}, \bibinfo{person}{Kenton Lee}, {and} \bibinfo{person}{Kristina
  Toutanova}.} \bibinfo{year}{2018}\natexlab{}.
\newblock \showarticletitle{Bert: Pre-training of deep bidirectional
  transformers for language understanding}.
\newblock \bibinfo{journal}{\emph{arXiv preprint arXiv:1810.04805}}
  (\bibinfo{year}{2018}).
\newblock


\bibitem[\protect\citeauthoryear{Fiorini, Canese, Starchenko, Kireev, Kim,
  Miller, Osipov, Kholodov, Ismagilov, Mohan, et~al\mbox{.}}{Fiorini
  et~al\mbox{.}}{2018a}]%
        {fiorini2018best}
\bibfield{author}{\bibinfo{person}{Nicolas Fiorini}, \bibinfo{person}{Kathi
  Canese}, \bibinfo{person}{Grisha Starchenko}, \bibinfo{person}{Evgeny
  Kireev}, \bibinfo{person}{Won Kim}, \bibinfo{person}{Vadim Miller},
  \bibinfo{person}{Maxim Osipov}, \bibinfo{person}{Michael Kholodov},
  \bibinfo{person}{Rafis Ismagilov}, \bibinfo{person}{Sunil Mohan},
  {et~al\mbox{.}}} \bibinfo{year}{2018}\natexlab{a}.
\newblock \showarticletitle{Best Match: new relevance search for PubMed}.
\newblock \bibinfo{journal}{\emph{PLoS biology}} \bibinfo{volume}{16},
  \bibinfo{number}{8} (\bibinfo{year}{2018}), \bibinfo{pages}{e2005343}.
\newblock


\bibitem[\protect\citeauthoryear{Fiorini, Leaman, Lipman, and Lu}{Fiorini
  et~al\mbox{.}}{2018b}]%
        {fiorini2018user}
\bibfield{author}{\bibinfo{person}{Nicolas Fiorini}, \bibinfo{person}{Robert
  Leaman}, \bibinfo{person}{David~J Lipman}, {and} \bibinfo{person}{Zhiyong
  Lu}.} \bibinfo{year}{2018}\natexlab{b}.
\newblock \showarticletitle{How user intelligence is improving PubMed}.
\newblock \bibinfo{journal}{\emph{Nature biotechnology}} \bibinfo{volume}{36},
  \bibinfo{number}{10} (\bibinfo{year}{2018}), \bibinfo{pages}{937}.
\newblock


\bibitem[\protect\citeauthoryear{Fiorini, Lipman, and Lu}{Fiorini
  et~al\mbox{.}}{2017}]%
        {fiorini2017cutting}
\bibfield{author}{\bibinfo{person}{Nicolas Fiorini}, \bibinfo{person}{David~J
  Lipman}, {and} \bibinfo{person}{Zhiyong Lu}.}
  \bibinfo{year}{2017}\natexlab{}.
\newblock \showarticletitle{Cutting edge: towards PubMed 2.0}.
\newblock \bibinfo{journal}{\emph{Elife}}  \bibinfo{volume}{6}
  (\bibinfo{year}{2017}), \bibinfo{pages}{e28801}.
\newblock


\bibitem[\protect\citeauthoryear{Gao, Pantel, Gamon, He, and Deng}{Gao
  et~al\mbox{.}}{2014}]%
        {gao2014modeling}
\bibfield{author}{\bibinfo{person}{Jianfeng Gao}, \bibinfo{person}{Patrick
  Pantel}, \bibinfo{person}{Michael Gamon}, \bibinfo{person}{Xiaodong He},
  {and} \bibinfo{person}{Li Deng}.} \bibinfo{year}{2014}\natexlab{}.
\newblock \showarticletitle{Modeling interestingness with deep neural
  networks}. In \bibinfo{booktitle}{\emph{Proceedings of the EMNLP}}.
  \bibinfo{pages}{2--13}.
\newblock


\bibitem[\protect\citeauthoryear{Goodwin and Harabagiu}{Goodwin and
  Harabagiu}{2016}]%
        {goodwin2016medical}
\bibfield{author}{\bibinfo{person}{Travis~R Goodwin} {and}
  \bibinfo{person}{Sanda~M Harabagiu}.} \bibinfo{year}{2016}\natexlab{}.
\newblock \showarticletitle{Medical question answering for clinical decision
  support}. In \bibinfo{booktitle}{\emph{CIKM}}. \bibinfo{pages}{297--306}.
\newblock


\bibitem[\protect\citeauthoryear{Guo, Fan, Ai, and Croft}{Guo
  et~al\mbox{.}}{2016}]%
        {guo2016deep}
\bibfield{author}{\bibinfo{person}{Jiafeng Guo}, \bibinfo{person}{Yixing Fan},
  \bibinfo{person}{Qingyao Ai}, {and} \bibinfo{person}{W~Bruce Croft}.}
  \bibinfo{year}{2016}\natexlab{}.
\newblock \showarticletitle{A deep relevance matching model for ad-hoc
  retrieval}. In \bibinfo{booktitle}{\emph{Proceedings of the CIKM}}.
  \bibinfo{pages}{55--64}.
\newblock


\bibitem[\protect\citeauthoryear{He, Zhang, Ren, and Sun}{He
  et~al\mbox{.}}{2015}]%
        {He:2015:DDR:2919332.2919814}
\bibfield{author}{\bibinfo{person}{Kaiming He}, \bibinfo{person}{Xiangyu
  Zhang}, \bibinfo{person}{Shaoqing Ren}, {and} \bibinfo{person}{Jian Sun}.}
  \bibinfo{year}{2015}\natexlab{}.
\newblock \showarticletitle{Delving Deep into Rectifiers: Surpassing
  Human-Level Performance on ImageNet Classification}. In
  \bibinfo{booktitle}{\emph{ICCV}}. \bibinfo{publisher}{IEEE Computer Society},
  \bibinfo{address}{Washington, DC, USA}, \bibinfo{pages}{1026--1034}.
\newblock


\bibitem[\protect\citeauthoryear{Hersh, Price, and Donohoe}{Hersh
  et~al\mbox{.}}{2000}]%
        {hersh2000assessing}
\bibfield{author}{\bibinfo{person}{William Hersh}, \bibinfo{person}{Susan
  Price}, {and} \bibinfo{person}{Larry Donohoe}.}
  \bibinfo{year}{2000}\natexlab{}.
\newblock \showarticletitle{Assessing thesaurus-based query expansion using the
  UMLS Metathesaurus.}. In \bibinfo{booktitle}{\emph{AMIA}}. American Medical
  Informatics Association, \bibinfo{pages}{344}.
\newblock


\bibitem[\protect\citeauthoryear{Hochreiter and Schmidhuber}{Hochreiter and
  Schmidhuber}{1997}]%
        {hochreiter1997long}
\bibfield{author}{\bibinfo{person}{Sepp Hochreiter} {and}
  \bibinfo{person}{J{\"u}rgen Schmidhuber}.} \bibinfo{year}{1997}\natexlab{}.
\newblock \showarticletitle{Long short-term memory}.
\newblock \bibinfo{journal}{\emph{Neural computation}} \bibinfo{volume}{9},
  \bibinfo{number}{8} (\bibinfo{year}{1997}), \bibinfo{pages}{1735--1780}.
\newblock


\bibitem[\protect\citeauthoryear{Hu, Lu, Li, and Chen}{Hu
  et~al\mbox{.}}{2014}]%
        {hu2014convolutional}
\bibfield{author}{\bibinfo{person}{Baotian Hu}, \bibinfo{person}{Zhengdong Lu},
  \bibinfo{person}{Hang Li}, {and} \bibinfo{person}{Qingcai Chen}.}
  \bibinfo{year}{2014}\natexlab{}.
\newblock \showarticletitle{Convolutional neural network architectures for
  matching natural language sentences}. In \bibinfo{booktitle}{\emph{NIPS}}.
  \bibinfo{pages}{2042--2050}.
\newblock


\bibitem[\protect\citeauthoryear{Huang, He, Gao, Deng, Acero, and Heck}{Huang
  et~al\mbox{.}}{2013}]%
        {huang2013learning}
\bibfield{author}{\bibinfo{person}{Po-Sen Huang}, \bibinfo{person}{Xiaodong
  He}, \bibinfo{person}{Jianfeng Gao}, \bibinfo{person}{Li Deng},
  \bibinfo{person}{Alex Acero}, {and} \bibinfo{person}{Larry Heck}.}
  \bibinfo{year}{2013}\natexlab{}.
\newblock \showarticletitle{Learning deep structured semantic models for web
  search using clickthrough data}. In \bibinfo{booktitle}{\emph{Proceedings of
  the CIKM}}. \bibinfo{pages}{2333--2338}.
\newblock


\bibitem[\protect\citeauthoryear{Hui, Yates, Berberich, and de~Melo}{Hui
  et~al\mbox{.}}{2017}]%
        {hui2017pacrr}
\bibfield{author}{\bibinfo{person}{Kai Hui}, \bibinfo{person}{Andrew Yates},
  \bibinfo{person}{Klaus Berberich}, {and} \bibinfo{person}{Gerard de Melo}.}
  \bibinfo{year}{2017}\natexlab{}.
\newblock \showarticletitle{{PACRR}: A Position-Aware Neural {IR} Model for
  Relevance Matching}. In \bibinfo{booktitle}{\emph{Proceedings of the EMNLP}}.
  \bibinfo{publisher}{Association for Computational Linguistics},
  \bibinfo{address}{Copenhagen, Denmark}, \bibinfo{pages}{1049--1058}.
\newblock


\bibitem[\protect\citeauthoryear{J{\"a}rvelin and
  Kek{\"a}l{\"a}inen}{J{\"a}rvelin and Kek{\"a}l{\"a}inen}{2002}]%
        {jarvelin2002cumulated}
\bibfield{author}{\bibinfo{person}{Kalervo J{\"a}rvelin} {and}
  \bibinfo{person}{Jaana Kek{\"a}l{\"a}inen}.} \bibinfo{year}{2002}\natexlab{}.
\newblock \showarticletitle{Cumulated gain-based evaluation of IR techniques}.
\newblock \bibinfo{journal}{\emph{ACM Transactions on Information Systems}}
  \bibinfo{volume}{20}, \bibinfo{number}{4} (\bibinfo{year}{2002}),
  \bibinfo{pages}{422--446}.
\newblock


\bibitem[\protect\citeauthoryear{Jiang, Zhang, Li, Bendersky, Golbandi, and
  Najork}{Jiang et~al\mbox{.}}{2019}]%
        {jiang2019semantic}
\bibfield{author}{\bibinfo{person}{Jyun-Yu Jiang}, \bibinfo{person}{Mingyang
  Zhang}, \bibinfo{person}{Cheng Li}, \bibinfo{person}{Mike Bendersky},
  \bibinfo{person}{Nadav Golbandi}, {and} \bibinfo{person}{Marc Najork}.}
  \bibinfo{year}{2019}\natexlab{}.
\newblock \showarticletitle{Semantic Text Matching for Long-Form Documents}.
\newblock  (\bibinfo{year}{2019}).
\newblock


\bibitem[\protect\citeauthoryear{Kipf and Welling}{Kipf and Welling}{2016}]%
        {kipf2016semi}
\bibfield{author}{\bibinfo{person}{Thomas~N Kipf} {and} \bibinfo{person}{Max
  Welling}.} \bibinfo{year}{2016}\natexlab{}.
\newblock \showarticletitle{Semi-supervised classification with graph
  convolutional networks}. In \bibinfo{booktitle}{\emph{ICLR}}.
\newblock


\bibitem[\protect\citeauthoryear{Lu, Kim, and Wilbur}{Lu et~al\mbox{.}}{2009}]%
        {lu2009evaluation}
\bibfield{author}{\bibinfo{person}{Zhiyong Lu}, \bibinfo{person}{Won Kim},
  {and} \bibinfo{person}{W~John Wilbur}.} \bibinfo{year}{2009}\natexlab{}.
\newblock \showarticletitle{Evaluation of query expansion using MeSH in
  PubMed}.
\newblock \bibinfo{journal}{\emph{Information retrieval}} \bibinfo{volume}{12},
  \bibinfo{number}{1} (\bibinfo{year}{2009}), \bibinfo{pages}{69--80}.
\newblock


\bibitem[\protect\citeauthoryear{Lu and Li}{Lu and Li}{2013}]%
        {lu2013deep}
\bibfield{author}{\bibinfo{person}{Zhengdong Lu} {and} \bibinfo{person}{Hang
  Li}.} \bibinfo{year}{2013}\natexlab{}.
\newblock \showarticletitle{A deep architecture for matching short texts}. In
  \bibinfo{booktitle}{\emph{NIPS}}. \bibinfo{pages}{1367--1375}.
\newblock


\bibitem[\protect\citeauthoryear{Manning, Surdeanu, Bauer, Finkel, Bethard, and
  McClosky}{Manning et~al\mbox{.}}{2014}]%
        {manning2014ACL}
\bibfield{author}{\bibinfo{person}{Christopher~D. Manning},
  \bibinfo{person}{Mihai Surdeanu}, \bibinfo{person}{John Bauer},
  \bibinfo{person}{Jenny Finkel}, \bibinfo{person}{Steven~J. Bethard}, {and}
  \bibinfo{person}{David McClosky}.} \bibinfo{year}{2014}\natexlab{}.
\newblock \showarticletitle{The {Stanford} {CoreNLP} Natural Language
  Processing Toolkit}. In \bibinfo{booktitle}{\emph{ACL System
  Demonstrations}}. \bibinfo{pages}{55--60}.
\newblock


\bibitem[\protect\citeauthoryear{Mikolov, Sutskever, Chen, Corrado, and
  Dean}{Mikolov et~al\mbox{.}}{2013}]%
        {mikolov2013distributed}
\bibfield{author}{\bibinfo{person}{Tomas Mikolov}, \bibinfo{person}{Ilya
  Sutskever}, \bibinfo{person}{Kai Chen}, \bibinfo{person}{Greg~S Corrado},
  {and} \bibinfo{person}{Jeff Dean}.} \bibinfo{year}{2013}\natexlab{}.
\newblock \showarticletitle{Distributed representations of words and phrases
  and their compositionality}. In \bibinfo{booktitle}{\emph{NIPS}}.
  \bibinfo{pages}{3111--3119}.
\newblock


\bibitem[\protect\citeauthoryear{Mohan, Fiorini, Kim, and Lu}{Mohan
  et~al\mbox{.}}{2018}]%
        {mohan2018fast}
\bibfield{author}{\bibinfo{person}{Sunil Mohan}, \bibinfo{person}{Nicolas
  Fiorini}, \bibinfo{person}{Sun Kim}, {and} \bibinfo{person}{Zhiyong Lu}.}
  \bibinfo{year}{2018}\natexlab{}.
\newblock \showarticletitle{A Fast Deep Learning Model for Textual Relevance in
  Biomedical Information Retrieval}. In \bibinfo{booktitle}{\emph{Proceedings
  of the WWW}}. \bibinfo{pages}{77--86}.
\newblock


\bibitem[\protect\citeauthoryear{Pennington, Socher, and Manning}{Pennington
  et~al\mbox{.}}{2014}]%
        {pennington2014glove}
\bibfield{author}{\bibinfo{person}{Jeffrey Pennington},
  \bibinfo{person}{Richard Socher}, {and} \bibinfo{person}{Christopher
  Manning}.} \bibinfo{year}{2014}\natexlab{}.
\newblock \showarticletitle{Glove: Global vectors for word representation}. In
  \bibinfo{booktitle}{\emph{Proceedings of the EMNLP}}.
  \bibinfo{pages}{1532--1543}.
\newblock


\bibitem[\protect\citeauthoryear{Qian, Santus, Jin, Guo, and Barzilay}{Qian
  et~al\mbox{.}}{2018}]%
        {qian2018graphie}
\bibfield{author}{\bibinfo{person}{Yujie Qian}, \bibinfo{person}{Enrico
  Santus}, \bibinfo{person}{Zhijing Jin}, \bibinfo{person}{Jiang Guo}, {and}
  \bibinfo{person}{Regina Barzilay}.} \bibinfo{year}{2018}\natexlab{}.
\newblock \showarticletitle{GraphIE: A graph-based framework for information
  extraction}.
\newblock \bibinfo{journal}{\emph{arXiv preprint arXiv:1810.13083}}
  (\bibinfo{year}{2018}).
\newblock


\bibitem[\protect\citeauthoryear{Roberts, Demner-Fushman, Voorhees, Hersh,
  Bedrick, Lazar, and Pant}{Roberts et~al\mbox{.}}{2017}]%
        {roberts2017overview}
\bibfield{author}{\bibinfo{person}{Kirk Roberts}, \bibinfo{person}{Dina
  Demner-Fushman}, \bibinfo{person}{Ellen~M Voorhees},
  \bibinfo{person}{William~R Hersh}, \bibinfo{person}{Steven Bedrick},
  \bibinfo{person}{Alexander~J Lazar}, {and} \bibinfo{person}{Shubham Pant}.}
  \bibinfo{year}{2017}\natexlab{}.
\newblock \showarticletitle{Overview of the TREC 2017 precision medicine
  track}.
\newblock \bibinfo{journal}{\emph{NIST Special Publication}}
  (\bibinfo{year}{2017}), \bibinfo{pages}{500--324}.
\newblock


\bibitem[\protect\citeauthoryear{Robertson, Walker, Jones, Hancock-Beaulieu,
  Gatford, et~al\mbox{.}}{Robertson et~al\mbox{.}}{1995}]%
        {robertson1995okapi}
\bibfield{author}{\bibinfo{person}{Stephen~E Robertson}, \bibinfo{person}{Steve
  Walker}, \bibinfo{person}{Susan Jones}, \bibinfo{person}{Micheline~M
  Hancock-Beaulieu}, \bibinfo{person}{Mike Gatford}, {et~al\mbox{.}}}
  \bibinfo{year}{1995}\natexlab{}.
\newblock \showarticletitle{Okapi at TREC-3}.
\newblock \bibinfo{journal}{\emph{Nist Special Publication Sp}}
  \bibinfo{volume}{109} (\bibinfo{year}{1995}), \bibinfo{pages}{109}.
\newblock


\bibitem[\protect\citeauthoryear{Scells, Zuccon, Koopman, Deacon, Azzopardi,
  and Geva}{Scells et~al\mbox{.}}{2017}]%
        {scells2017integrating}
\bibfield{author}{\bibinfo{person}{Harrisen Scells}, \bibinfo{person}{Guido
  Zuccon}, \bibinfo{person}{Bevan Koopman}, \bibinfo{person}{Anthony Deacon},
  \bibinfo{person}{Leif Azzopardi}, {and} \bibinfo{person}{Shlomo Geva}.}
  \bibinfo{year}{2017}\natexlab{}.
\newblock \showarticletitle{Integrating the framing of clinical questions via
  PICO into the retrieval of medical literature for systematic reviews}. In
  \bibinfo{booktitle}{\emph{Proceedings of the CIKM}}.
  \bibinfo{pages}{2291--2294}.
\newblock


\bibitem[\protect\citeauthoryear{Soldaini, Yates, and Goharian}{Soldaini
  et~al\mbox{.}}{2017}]%
        {Soldaini2017cikm}
\bibfield{author}{\bibinfo{person}{Luca Soldaini}, \bibinfo{person}{Andrew
  Yates}, {and} \bibinfo{person}{Nazli Goharian}.}
  \bibinfo{year}{2017}\natexlab{}.
\newblock \showarticletitle{Denoising Clinical Notes for Medical Literature
  Retrieval with Convolutional Neural Model}. In
  \bibinfo{booktitle}{\emph{CIKM}}. \bibinfo{pages}{2307--2310}.
\newblock


\bibitem[\protect\citeauthoryear{Song, Zhang, Wang, and Gildea}{Song
  et~al\mbox{.}}{2018}]%
        {song2018n}
\bibfield{author}{\bibinfo{person}{Linfeng Song}, \bibinfo{person}{Yue Zhang},
  \bibinfo{person}{Zhiguo Wang}, {and} \bibinfo{person}{Daniel Gildea}.}
  \bibinfo{year}{2018}\natexlab{}.
\newblock \showarticletitle{N-ary relation extraction using graph state lstm}.
\newblock \bibinfo{journal}{\emph{arXiv preprint arXiv:1808.09101}}
  (\bibinfo{year}{2018}).
\newblock


\bibitem[\protect\citeauthoryear{Wei, Kao, and Lu}{Wei et~al\mbox{.}}{2013}]%
        {wei2013pubtator}
\bibfield{author}{\bibinfo{person}{Chih-Hsuan Wei}, \bibinfo{person}{Hung-Yu
  Kao}, {and} \bibinfo{person}{Zhiyong Lu}.} \bibinfo{year}{2013}\natexlab{}.
\newblock \showarticletitle{PubTator: a web-based text mining tool for
  assisting biocuration}.
\newblock \bibinfo{journal}{\emph{Nucleic acids research}}
  \bibinfo{volume}{41}, \bibinfo{number}{W1} (\bibinfo{year}{2013}),
  \bibinfo{pages}{518--522}.
\newblock


\bibitem[\protect\citeauthoryear{Xiong and Callan}{Xiong and Callan}{2015a}]%
        {xiong2015esdrank}
\bibfield{author}{\bibinfo{person}{Chenyan Xiong} {and} \bibinfo{person}{Jamie
  Callan}.} \bibinfo{year}{2015}\natexlab{a}.
\newblock \showarticletitle{Esdrank: Connecting query and documents through
  external semi-structured data}. In \bibinfo{booktitle}{\emph{CIKM}}. ACM,
  \bibinfo{pages}{951--960}.
\newblock


\bibitem[\protect\citeauthoryear{Xiong and Callan}{Xiong and Callan}{2015b}]%
        {Xiong2015}
\bibfield{author}{\bibinfo{person}{Chenyan Xiong} {and} \bibinfo{person}{Jamie
  Callan}.} \bibinfo{year}{2015}\natexlab{b}.
\newblock \showarticletitle{Query Expansion with Freebase}. In
  \bibinfo{booktitle}{\emph{Proceedings of the 2015 International Conference on
  The Theory of Information Retrieval}}. \bibinfo{pages}{111--120}.
\newblock


\bibitem[\protect\citeauthoryear{Xiong, Dai, Callan, Liu, and Power}{Xiong
  et~al\mbox{.}}{2017a}]%
        {xiong2017end}
\bibfield{author}{\bibinfo{person}{Chenyan Xiong}, \bibinfo{person}{Zhuyun
  Dai}, \bibinfo{person}{Jamie Callan}, \bibinfo{person}{Zhiyuan Liu}, {and}
  \bibinfo{person}{Russell Power}.} \bibinfo{year}{2017}\natexlab{a}.
\newblock \showarticletitle{End-to-end neural ad-hoc ranking with kernel
  pooling}. In \bibinfo{booktitle}{\emph{Proceedings of the SIGIR}}. ACM,
  \bibinfo{pages}{55--64}.
\newblock


\bibitem[\protect\citeauthoryear{Xiong, Power, and Callan}{Xiong
  et~al\mbox{.}}{2017b}]%
        {xiong2017explicit}
\bibfield{author}{\bibinfo{person}{Chenyan Xiong}, \bibinfo{person}{Russell
  Power}, {and} \bibinfo{person}{Jamie Callan}.}
  \bibinfo{year}{2017}\natexlab{b}.
\newblock \showarticletitle{Explicit semantic ranking for academic search via
  knowledge graph embedding}. In \bibinfo{booktitle}{\emph{WWW}}. International
  World Wide Web Conferences Steering Committee, \bibinfo{pages}{1271--1279}.
\newblock


\bibitem[\protect\citeauthoryear{Zhao, Jiang, Liu, Qin, and Liu}{Zhao
  et~al\mbox{.}}{2018}]%
        {zhao2018causaltriad}
\bibfield{author}{\bibinfo{person}{Sendong Zhao}, \bibinfo{person}{Meng Jiang},
  \bibinfo{person}{Ming Liu}, \bibinfo{person}{Bing Qin}, {and}
  \bibinfo{person}{Ting Liu}.} \bibinfo{year}{2018}\natexlab{}.
\newblock \showarticletitle{CausalTriad: Toward Pseudo Causal Relation
  Discovery and Hypotheses Generation from Medical Text Data}. In
  \bibinfo{booktitle}{\emph{ACM BCB}}. ACM, \bibinfo{pages}{184--193}.
\newblock


\bibitem[\protect\citeauthoryear{Zhao, Liu, Zhao, and Wang}{Zhao
  et~al\mbox{.}}{2019}]%
        {zhao2018neural}
\bibfield{author}{\bibinfo{person}{Sendong Zhao}, \bibinfo{person}{Ting Liu},
  \bibinfo{person}{Sicheng Zhao}, {and} \bibinfo{person}{Fei Wang}.}
  \bibinfo{year}{2019}\natexlab{}.
\newblock \showarticletitle{A Neural Multi-Task Learning Framework to Jointly
  Model Medical Named Entity Recognition and Normalization}. In
  \bibinfo{booktitle}{\emph{AAAI}}, Vol.~\bibinfo{volume}{33}.
  \bibinfo{pages}{817--824}.
\newblock


\bibitem[\protect\citeauthoryear{Zhao, Wang, Massung, Qin, Liu, Wang, and
  Zhai}{Zhao et~al\mbox{.}}{2017}]%
        {zhao2017constructing}
\bibfield{author}{\bibinfo{person}{Sendong Zhao}, \bibinfo{person}{Quan Wang},
  \bibinfo{person}{Sean Massung}, \bibinfo{person}{Bing Qin},
  \bibinfo{person}{Ting Liu}, \bibinfo{person}{Bin Wang}, {and}
  \bibinfo{person}{ChengXiang Zhai}.} \bibinfo{year}{2017}\natexlab{}.
\newblock \showarticletitle{Constructing and embedding abstract event causality
  networks from text snippets}. In \bibinfo{booktitle}{\emph{WSDM}}. ACM,
  \bibinfo{pages}{335--344}.
\newblock


\bibitem[\protect\citeauthoryear{Zheng and Wan}{Zheng and Wan}{2016}]%
        {Zheng2016graph}
\bibfield{author}{\bibinfo{person}{Ziwei Zheng} {and} \bibinfo{person}{Xiaojun
  Wan}.} \bibinfo{year}{2016}\natexlab{}.
\newblock \showarticletitle{Graph-Based Multi-Modality Learning for Clinical
  Decision Support}. In \bibinfo{booktitle}{\emph{Proceedings of the CIKM}}.
  \bibinfo{pages}{1945--1948}.
\newblock


\bibitem[\protect\citeauthoryear{Zhu, Elemento, Pathak, and Wang}{Zhu
  et~al\mbox{.}}{2018}]%
        {zhu2018drug}
\bibfield{author}{\bibinfo{person}{Yongjun Zhu}, \bibinfo{person}{Olivier
  Elemento}, \bibinfo{person}{Jyotishman Pathak}, {and} \bibinfo{person}{Fei
  Wang}.} \bibinfo{year}{2018}\natexlab{}.
\newblock \showarticletitle{Drug knowledge bases and their applications in
  biomedical informatics research}.
\newblock \bibinfo{journal}{\emph{Briefings in bioinformatics}}
  (\bibinfo{year}{2018}).
\newblock


\end{thebibliography}

\end{document}